\documentstyle[preprint,aps,prc,eqsecnum]{revtex}
\begin{document}
\draft
\preprint{RUB-MEP-163/95}
\tighten
\title{Does The 3N-Force Have A Hard Core?}
\author{J. A. Eden and M. F. Gari}
\address{ Institut f\"ur Theoretische Physik\\
	  Ruhr Universit\"at Bochum, D-44780 Bochum, Germany\\
	  {\rm email: jamie@deuteron.tp2.ruhr-uni-bochum.de}}
\date{17 October 1995; Revised 8 January 1996}
\maketitle
\begin{abstract}
The meson-nucleon dynamics that generates the hard core of the RuhrPot 
two-nucleon interaction is shown to vanish in the irreducible 3N force.
This result indicates a small 3N force dominated by conventional
light meson-exchange dynamics and holds for an arbitrary meson-theoretic 
Lagrangian. The resulting RuhrPot 3N force is defined in the appendix.
A completely different result is expected when the
Tamm-Dancoff/Bloch-Horowitz procedure is used to define the NN and 
3N potentials. In that approach, (e.g. full Bonn potential) both the
NN {\it and} 3N potentials contain non-vanishing contributions from the 
coherent sum of meson-recoil dynamics and the possibility of a 
large hard core requiring explicit calculation cannot be ruled out.
\end{abstract}
%
%%%%%%%%%%%%%%%%%%%%%%%%%%%%%%%%%%% Chapter 1 %%%%%%%%%%%%%%%%%%%%%%%%%%%%%%%%%%
\narrowtext
\section{Introduction}\label{sec1}
Existing descriptions of the 3-nucleon force \cite{FR92} essentially fall into 
two categories. The first is the semi-phenomenological approach adopted for the 
calculation of the Tucson-Melbourne force \cite{TM1,TM2,TM3,TM4,TM5,TM6,TM7},
where the $\pi\pi$, $\pi\rho$ and $\rho\rho$ contributions to the 3N 
interaction are fixed by subtracting the forward-propagating Born amplitudes 
from $\pi$N-scattering, photo-pion production and photon-scattering data and 
the off-shell extrapolation that is necessary for 3-body applications is 
obtained from PCAC and current algebra. There is little doubt that at $Q^2$=0, 
this force can be regarded as essentially exact. The second approach, which has
been adopted for the calculation of the Brazilian \cite{BR1,BR2,BR3,BR4,BR5} 
and RuhrPot 3N-forces is to make use of a particular meson-theoretic model and 
directly calculate the leading-order contributions. We illustrate the two 
schemes in Fig~1.

One important question that remains to be addressed is this: {\it `Does the 
3N-interaction have a hard core like that found in the NN-interaction?'}. 
To address this question, we will first recall the dynamical origins of the 
hard core in the NN interaction, and then ask if such dynamics can produce a 
corresponding hard core in the 3-body system.

About 5 years ago it was shown \cite{RP91} that the hard core of the NN-nucleon
interaction arises naturally in a boson-exchange model when the dynamics is no 
longer truncated to include the exchange of only a few light mesons. 
In particular, the RuhrPot NN-interaction \cite{RP91,RP94} introduces a closure
approximation to incorporate the additional $(J^\pi;T)$-exchange dynamics that 
is required by completeness. Conventional light-meson-exchange dynamics still 
contributes to the NN-interaction at long distances, but the hard core region 
is now completely dominated by the additional {\it contact} contributions. 
We illustrate this two-phase approach in Fig.~2.

This natural separation of the long- and short-range NN-interaction dynamics has
proven critical in resolving a number of problems in existing boson-exchange 
potentials (BEPs). For example, when the contact interactions are included to 
describe the hard core of the NN interaction, it is no longer necessary to 
adopt the artificial NN-meson cut-offs that are required \cite{AT89} in 
conventional boson-exchange models. Instead, the RuhrPot NN interaction 
and  associated exchange currents \cite{JE95a,JE95b}
use self-consistently calculated form factors \cite{KG83,JF95} that possess an 
asymptotic $Q^2$ dependence which is consistent \cite{GK85} with perturbative 
QCD \cite{BR81}. In addition, the vector-meson couplings satisfy the 
${\rm SU(3)}_{\rm F}$ prediction of 
$g^2_{{\rm NN}\omega}/g^2_{{\rm NN}\rho}\sim 9$, which can be compared to 
values near 27 that are typical \cite{Bonn} of conventional boson-exchange 
potentials.

So how do the contact interactions effect the irreducible 3N force? 
What happens to the 3-body force when two or three of the nucleons begin 
to overlap? Our answers to these questions are organized as follows.
In section~2, we summarize the Tamm-Dancoff/Bloch-Horowitz and
unitary transformation projection formalisms that define the one- and 
two-boson-exchange interactions in the A-body system. 
In section~3 we use these results to recall how the hard core of the
NN~interaction arises naturally when the exchange dynamics is no longer 
arbitrarily truncated to include only the first few light mesons.
In section~4 we carry these ideas into the 3N system to investigate the
existence of a possible hard core in the irreducible 3N interaction.
Within the unitary transformation approach, we find that {\it the class 
of $(J^\pi;T)$-exchange processes generating the hard core repulsion in 
the RuhrPot NN interaction vanish in the non-relativistic limit of the 
irreducible 3N interaction.} This result is not confined to any particular 
meson-theoretic model and it eliminates the most likely source of a hard 
core in the 3N interaction. 
However, a completely different result is expected when the
Tamm-Dancoff/Bloch-Horowitz procedure is used to define the NN and 
3N potentials. In that approach, (e.g. full Bonn potential) both the
NN {\it and} 3N potentials contain non-vanishing contributions from the 
coherent sum of meson-recoil dynamics and the possibility of a 
large hard core requiring explicit calculation cannot be ruled out.
Our conclusions are presented in section~5.

%%%%%%%%%%%%%%%%%%%%%%%%%%%%%%%%%%% Chapter 2 %%%%%%%%%%%%%%%%%%%%%%%%%%%%%%%%%
\section{Formalism}\label{sec2}
The energy eigen-value problem for an arbitrary interacting meson-baryon system
is given by, 
\begin{equation}\label{2.1}
H\vert\Psi\rangle=(H_0+H_I)\vert\Psi\rangle= E_i\vert\Psi\rangle
\end{equation}
where the total Hamiltonian $H$ is separated into free and interacting parts 
$H_0$ and $H_I$, and $\vert\Psi\rangle$ denotes the complete meson-baryon state with
energy $E_i$. This provides a relativistic time-ordered framework, rather than 
a manifestly covariant one.
A rigorous solution of eq.~(\ref{2.1}) in it's present form is impossible 
because the wave functions contain not only nucleon degrees of freedom, 
but explicit meson, resonance and anti-nucleonic degrees of freedom as well.
However, as is well known, the problem can be reduced to a tractable form by 
partitioning the total Hilbert space into two parts,
\begin{eqnarray}\label{2.2}
 {\cal H}_{\eta}    &=& \Bigl\{ \vert {\rm N}^{(+)}\rangle, 
                              \vert {\rm N}^{(+)}{\rm N}^{(+)}\rangle, 
                              ... \Bigr\} 
\nonumber \\
 {\cal H}_{\lambda} &=& \Bigl\{ {\rm Everything~~else} \Bigr\}
\end{eqnarray}
so that ${\cal H}_{\eta}$ is the Hilbert sub-space consisting only of 
the positive-frequency parts of the nucleon state vectors, 
and ${\cal H}_{\lambda}$ contains everything else, e.g. $\vert\Delta\rangle$, 
$\vert $N$\pi\rangle$,  $\vert $N$\Delta\pi\rho\rangle$ etc. 
The total wave function can now be expressed as,
\begin{eqnarray}\label{2.3}
\vert \Psi \rangle  &=& \eta \vert \Psi \rangle + \lambda \vert \Psi \rangle
                    = \vert \psi \rangle + \vert \phi \rangle
\end{eqnarray}
where we have introduced projection operators satisfying the conventional
algebra $\eta^2=\eta$, $\lambda^2=\lambda$, $\eta\lambda=\lambda\eta=0$ and 
$\eta+\lambda=1$ to obtain 
$\vert\psi\rangle=\eta\vert\Psi\rangle\in{\cal H}_{\eta}$ 
as a purely nucleonic state and 
$\vert\phi\rangle=\lambda\vert\Psi\rangle\in{\cal H}_{\lambda}$ 
as a state whose description requires explicit meson- and/or resonance- and/or 
negative-frequency degrees of freedom.
We have not given explicit expressions for $\eta$ and 
$\lambda$ - nor will we need to since 
$\vert\psi\rangle$=$\eta\vert\psi\rangle$ and
$\vert\phi\rangle$=$\lambda\vert\phi\rangle$.
The point is that the energy eigen value problem can now be written as,
\begin{mathletters}\label{2.4}
\begin{eqnarray}
E_i \vert \psi \rangle  &=&
\eta H\eta       \vert \psi \rangle  + \eta H\lambda    \vert \phi \rangle 
\\
E_i \vert \phi \rangle  &=&
\lambda H\eta    \vert \psi \rangle  + \lambda H\lambda \vert \phi \rangle 
\end{eqnarray}
\end{mathletters}
There are a number of different (but equivalent \cite{FST,SO54}) ways in which 
eq.~(\ref{2.4}) can be reduced when $\vert \phi \rangle$ vanishes in the 
observable states. One seeks to derive an {\it effective} Hamiltonian 
$H_{\rm eff}$ (or an effective interaction $V_{\rm eff}$) which satisfies,
\begin{equation}\label{2.5}
  H_{\rm eff} \vert \psi \rangle = [ H_0 + V_{\rm eff} ] \vert \psi \rangle 
= E_i \vert \psi \rangle 
\end{equation}
so that the matrix elements of the Hamiltonian between interacting meson-baryon
wave functions $\vert \Psi \rangle$, as required in eq.~(\ref{2.1}), can be 
computed as an effective Hamiltonian between conventional nucleonic 
wave functions $\vert \psi \rangle$.

Perhaps the most commonly adopted reduction scheme is found in the Tamm-Dancoff 
approximation \cite{TDA}, or in one of the many equivalent schemes like that 
due to Bloch and Horowitz \cite{BH58}.
Here one begins by noting that the free-energy Hamiltonian cannot cause 
transitions between the Hilbert sub-spaces, so $\lambda H_0\eta$=0 can be used 
to reduce eq.~(\ref{2.4}b) to, 
\begin{equation}\label{2.6}
(E_i  - H_0 ) \lambda \vert \phi \rangle  =
  \lambda H_I [\eta    \vert \psi \rangle  +  \lambda \vert \phi \rangle ]
\end{equation}
Collecting the $\lambda\vert\phi\rangle$ terms together gives \cite{SO54} 
Sawada's result,
\begin{equation} \label{2.7}
\lambda\vert\phi\rangle = 
{1\over [1 - {\lambda\over(E_i - H_0)}H_I]} 
             {\lambda\over(E_i - H_0)}H_I \eta\vert\psi\rangle.
\end{equation}
After inserting this into Eq.~(\ref{2.4}a) and noting that $\eta H_I\eta$=0, 
a comparison with eq.~(\ref{2.5}) shows that
\begin{equation} \label{2.8}
V_{\rm eff} =
H_I {1\over [1 - {\lambda\over(E_i - H_0)}H_I]} {\lambda\over(E_i - H_0)}H_I 
\qquad
\end{equation}
Equivalently, we can recognize eq.~(\ref{2.6}) as a recursive definition of 
$\vert\phi\rangle$, so that,
\begin{equation} \label{2.9}
\lambda\vert\phi\rangle = {\lambda\over(E_i - H_0)}R\vert\psi\rangle,
\qquad
R = H_I + H_I {\lambda\over (E_i-H_0)} R
\end{equation}
and
\begin{equation}\label{2.10}
V_{\rm eff}  = H_I {\lambda\over (E_i  - H_0 )} R  
\end{equation}
Expanding the interaction Hamiltonian in powers of the strong coupling,
\begin{equation}\label{2.11}
H_I = \sum_{i=1}^{\infty} H_i
\end{equation}
allows for the calculation of the effective interaction to any desired order.
In particular, the one-boson exchange potential in the A-nucleon system is 
given by
\begin{eqnarray}\label{2.12}
 V^{(2?)}_{\rm eff} &=&
  H_1 {\lambda\over (E_i  - H_0 )} H_1
\end{eqnarray}
and the corresponding two-boson exchange potential is,
\begin{eqnarray}\label{2.13}
 V^{(4?)}_{\rm eff} &=&
  H_1 {\lambda\over (E_i-H_0 )} H_1 {\lambda\over (E_i-H_0 )} 
                                H_1 {\lambda\over (E_i  - H_0 )} H_1 
\nonumber \\ && 
+ H_2 {\lambda\over (E_i-H_0 )} H_1 {\lambda\over (E_i-H_0 )} H_1
\nonumber \\ && 
+ H_1 {\lambda\over (E_i-H_0 )} H_2 {\lambda\over (E_i-H_0 )} H_1
\nonumber \\ && 
+ H_1 {\lambda\over (E_i-H_0 )} H_1 {\lambda\over (E_i-H_0 )} H_2
\nonumber \\ && 
+ H_2 {\lambda\over (E_i-H_0 )} H_2
\nonumber \\ && 
\end{eqnarray}
We have labeled the order of the interactions in eqs.~(\ref{2.12}) and (\ref{2.13}) 
with question marks since these expressions are obtained from
eq.~(\ref{2.10}) by expanding the numerator to definite order, but retaining
the full (infinite order) energy dependence in the dominator.

We now appear to have effective interactions for usage in eq.~(\ref{2.5}).
Indeed, if we set $H_2$=0 in eqs.~(\ref{2.12}) and (\ref{2.13}), 
we obtain the OBEP and TBEP interactions used in the full Bonn potential.
However, it is easy to see that the Hermiticity of these effective interactions
is destroyed by an explicit dependence on the full initial-state energy $E_i$. 
In addition, from eq.~(\ref{2.3}) and (\ref{2.9}) we observe that the wave 
functions are not orthonormal but instead satisfy
\begin{equation}\label{2.14}
\delta_{fi} =
\langle\psi_f\vert \left[ \openone + R^\dagger{\lambda\over E_f-H_0}
                          {\lambda\over E_i-H_0} R\right] 
\vert \psi_i\rangle
\end{equation}
As such, the orthonormality condition depends on the order at which we truncate 
the interaction.  These are certainly unwanted complications that will invite 
dubious approximation.

The problems  associated with the Tamm-Dancoff approach can be removed 
\cite{FST} by expanding the energy $E$ in the same way as we expanded the 
Hamiltonian in eq.~(\ref{2.11}). It is then possible to obtain an interaction 
to a definite order and to calculate a renormalization condition for the 
wave functions.  An equivalent solution, which was developed by 
Okubo \cite{SO54} and which we will refer to as the {\it unitary transformation
method}, proceeds by rewriting eq.~(\ref{2.4}) in matrix form as,
\begin{equation}\label{2.15}
E_i \pmatrix{ \vert \psi \rangle \cr \vert \phi \rangle }
 =
\pmatrix{ \eta H\eta  &    \eta H\lambda \cr
         \lambda H\eta  & \lambda H\lambda}
\pmatrix{ \vert \psi \rangle \cr \vert \phi \rangle }
\end{equation}
and introducing new states $\vert\chi\rangle$ and $\vert\varphi\rangle$
through a unitary transformation,
\begin{equation}\label{2.16}
   \pmatrix{\vert\psi\rangle\cr\vert\phi\rangle\cr}
=U \pmatrix{\vert\chi\rangle\cr\vert\varphi\rangle}
\end{equation}
In terms of the new states, the energy eigenvalue problem reads,
\begin{equation}\label{2.17}
U^\dagger  H U \pmatrix{\vert\chi\rangle\cr\vert\varphi\rangle}
=   E_i \pmatrix{\vert\chi\rangle\cr\vert\varphi\rangle}
\end{equation}
Unlike the Tamm-Dancoff reduction, where we found that eqs.~(\ref{2.7}) and
(\ref{2.9}) related $\vert\psi\rangle$ and $\vert\phi\rangle$, we can now
choose $U$ such that it is unitary and ensures $U^\dagger H U$ is diagonal, 
so that $\vert\chi\rangle$ and $\vert\varphi\rangle$ are completely decoupled 
orthonormal states.  In particular, for
\begin{equation}\label{2.18}
U=\pmatrix{
 \eta(1+A^\dagger A)^{-{1\over 2}}\eta &
-\eta A^\dagger(1+AA^\dagger)^{-{1\over2}}\lambda \cr
 \lambda A (1+A^\dagger A)^{-{1\over 2}} \eta & 
 \lambda (1+AA^\dagger)^{-{1\over2}}\lambda \cr}
\end{equation}
a short calculation shows that the non-diagonal elements of eq.~(\ref{2.17}) 
vanish for any $A$ which satisfies,
\begin{equation}\label{2.19}
\lambda \left( H+ \left[H,A\right] -AHA \right) \eta =0 
\end{equation}
We will derive the operators $A$ that satisfy eq.~(\ref{2.19}) in a moment.
When $\vert\varphi\rangle$ vanishes in the observable states, the energy 
eigen value problem reduces to the conventional Schr\"odinger equation 
\begin{equation}\label{2.20}
  \langle \Psi \vert  H           \vert \Psi \rangle 
= \langle \chi \vert  H_{\rm eff} \vert \chi \rangle 
= \langle \chi \vert E_i \vert \chi \rangle 
\end{equation}
with the Hermitean effective interaction,
\begin{equation}\label{2.21}
H_{\rm eff}=\eta(1+A^\dagger A)^{-{1\over 2}}\eta (1+A^\dagger) H (1+A) \eta
            (1+A^\dagger A)^{-{1\over 2}}\eta
\end{equation}
The complete meson-baryon wave functions $\vert \Psi \rangle$ 
are then described in terms of the operators $A$ and the nucleonic
wave functions $\vert \chi \rangle $,
\begin{equation}\label{2.22}
\vert\Psi\rangle=(1+A)\eta(1+A^\dagger A)^{-{1\over 2}}\eta\vert\chi\rangle
\end{equation}
As already anticipated, we find $\vert\chi\rangle\in{\cal H}_{\eta}$  and
$\vert\varphi\rangle\in{\cal H}_{\lambda}$ preserve the orthonormality of 
$\vert\Psi\rangle$,
\widetext
\begin{mathletters}\label{2.23}
\begin{eqnarray}
\delta_{fi} = \langle\Psi_f\vert\Psi_i\rangle 
&=&
\langle \chi_f \vert \eta (1+A^\dagger A)^{-{1\over 2}} \eta (1+A^\dagger)(1+A) \eta
                  (1+A^\dagger A)^{-{1\over 2}} \eta \vert\chi_i\rangle 
=\langle\chi_f\vert \chi_i \rangle
\\
\delta_{fi} = \langle\Psi_f\vert\Psi_i\rangle 
&=&
\langle\varphi_f\vert \eta (1+AA^\dagger)^{-{1\over 2}} \eta 
    (1-A)(1-A^\dagger)\eta
    (1+AA^\dagger)^{-{1\over 2}} \eta \vert\varphi_i\rangle 
=\langle\varphi_f\vert \varphi_i \rangle
\end{eqnarray}
\end{mathletters}
\narrowtext
We next derive the operators $A$ that satisfy eq.~(\ref{2.19})
by expanding $A$ and the Hamiltonian to which we already know it is 
related, in powers $n$ of the coupling constant,
\begin{equation}\label{2.24}
H = H_0 + \sum_{n=1}^{\infty}  H_n,\qquad
A =\sum_{n=1}^{\infty}  A_n
\end{equation}
where  $A$=$\lambda A \eta$ shows that $A_0$=0.  Similarly, we note that 
$\eta H_I \eta$ = $\eta H_0 \lambda$ = $\lambda H_0 \eta$ =0,
so that eq.~(\ref{2.19}) becomes,
\begin{eqnarray}\label{2.25}
0 &=& \sum_{n=1}^{\infty} \lambda \Biggl[
      H_n +  [H_0\, , A_n]
      + \sum_{i=1}^{n-1} H_i A_{n-i}
\nonumber \\ &&
      -\sum_{i=1}^{n-2} \sum_{j=1}^{n-i-1} A_i H_j A_{n-i-j} \Biggr]\eta
\end{eqnarray}
We choose to further constrain $A$ by demanding eq.~(\ref{2.25}) is satisfied 
at each order of $n$. With $H_0\,\eta$=${\cal E}_i\,\eta$, where ${\cal E}_i$
is the free-particle energy of the initial state, we obtain,
\begin{eqnarray}\label{2.26}
({\cal E}_i-H_0) A_n &=& \lambda \Biggl[
   H_n 
+ \sum_{i=1}^{n-1} H_i A_{n-i}
\nonumber \\ &&
- \sum_{i=1}^{n-2}\,\, \sum_{j=1}^{n-i-1} A_i H_j A_{n-i-j}
\Biggr]\eta
\end{eqnarray}
This completes the definition of A. To calculate the OBEP and TBEP interactions 
in the A-body system, we require, 
\begin{mathletters}\label{2.27}
\begin{eqnarray}
A_1
&=& {\lambda\over {\cal E}_i-H_0} H_1 \eta  
\\
A_2
&=& {\lambda\over {\cal E}_i-H_0} H_2 \eta 
 +  {\lambda\over {\cal E}_i-H_0} H_1 {\lambda\over {\cal E}_i-H_0} H_1 \eta 
\\
A_3
&=& {\lambda\over {\cal E}_i-H_0} H_3 \eta  
 + {\lambda\over {\cal E}_i-H_0} H_1 {\lambda\over {\cal E}_i-H_0} H_2 \eta 
\nonumber \\ 
 &+& {\lambda\over {\cal E}_i-H_0} H_1 {\lambda\over {\cal E}_i-H_0} H_1
                                      {\lambda\over {\cal E}_i-H_0} H_1 \eta 
\nonumber \\
 &+& {\lambda\over {\cal E}_i-H_0} H_2 {\lambda\over {\cal E}_i-H_0} H_1 
\nonumber \\ 
 &-&  {\lambda\over {\cal E}_i-H_0}{\lambda\over {\cal E}_a-H_0} 
    H_1 \eta_a H_1 {\lambda\over {\cal E}_i-H_0} H_1  \eta 
\nonumber
\end{eqnarray}
\end{mathletters}
where ${\cal E}_a$ is the free-particle energy of an intermediate $\eta$-space 
state. The OBEP in the A-body system is then given by
\begin{eqnarray}\label{2.28}
&&V^{(2)}_{\rm eff}= \eta_f \Biggl\{
  H_1 \lambda {{1\over 2}({\cal E}_i+{\cal E}_f)-H_0 
  \over ({\cal E}_f-H_0)({\cal E}_i-H_0)} \lambda H_1 
\Biggr\}\eta_i 
\end{eqnarray}
and the corresponding TBEP is,
\widetext
\begin{eqnarray}\label{2.29}
&& V^{(4)}_{\rm eff} = \eta_f \Biggl\{
H_1 {\lambda\over {\cal E}_f-H_0} H_1 \lambda 
{{1\over 2}({\cal E}_f+{\cal E}_i)-H_0\over ({\cal E}_f-H_0)({\cal E}_i-H_0)} 
\lambda H_1 {\lambda\over {\cal E}_i-H_0} H_1 
\nonumber \\ &-&
{1\over 2} H_1 {\lambda\over {\cal E}_f-H_0} \Biggl[
  {\lambda\over E_a-H_0} H_1 \eta_a H_1 
+ H_1 \eta_a H_1 {\lambda\over E_a-H_0} 
\Biggr] {\lambda\over {\cal E}_i-H_0} H_1 
\nonumber \\ &+&
 H_2 \lambda {{1\over 2}({\cal E}_f+{\cal E}_i)-H_0 
       \over ({\cal E}_f-H_0)({\cal E}_i-H_0)} \lambda H_2
\nonumber \\ &+&
{1\over 2}({\cal E}_f-{\cal E}_i) H_1 {\lambda\over {\cal E}_f-H_0} \Biggl[
  H_1 \eta_a H_1 {\lambda \over (E_a-H_0)({\cal E}_f-H_0)} 
 +{\lambda\over({\cal E}_i-H_0)(E_a-H_0)} H_1 \eta_a H_1
\nonumber \\ && \qquad
-H_1 {\lambda\over {\cal E}_f-H_0} H_1       {\lambda\over {\cal E}_f-H_0} 
-{\lambda\over {\cal E}_i-H_0} H_1 {\lambda\over {\cal E}_i-H_0} H_1
\Biggr] {\lambda\over {\cal E}_i-H_0} H_1 
\nonumber \\ &-&
\left[ {{\cal E}_f+{\cal E}_i\over 8}-{E_a\over 2}\right]
           H_1 {\lambda\over(E_a-H_0)({\cal E}_f-H_0)} H_1 \eta_a
           H_1 {\lambda\over(E_a-H_0)({\cal E}_i-H_0)} H_1 
\nonumber \\ &+&
  H_2 {\lambda\over {\cal E}_f-H_0} H_1 {\lambda\over {\cal E}_i-H_0} H_1
+ H_1 {\lambda\over {\cal E}_f-H_0} H_2 {\lambda\over {\cal E}_i-H_0} H_1
+ H_1 {\lambda\over {\cal E}_f-H_0} H_1 {\lambda\over {\cal E}_i-H_0} H_2
\nonumber \\ &+&
{1\over 2}({\cal E}_f-{\cal E}_i) H_2{\lambda\over {\cal E}_f-H_0} 
\Bigl[ {\lambda\over {\cal E}_i-H_0} H_1 -H_1{\lambda\over {\cal E}_f-H_0} 
\Bigr] {\lambda\over {\cal E}_i-H_0} H_1 
\nonumber \\ &+&
{1\over 2}({\cal E}_f-{\cal E}_i)H_1{\lambda\over {\cal E}_f-H_0}
\Bigl[ {\lambda\over {\cal E}_i-H_0} H_2 -H_2{\lambda\over {\cal E}_f-H_0} 
\Bigr] {\lambda\over {\cal E}_i-H_0} H_1 
\nonumber \\ &+&
{1\over 2}({\cal E}_f-{\cal E}_i)H_1{\lambda\over {\cal E}_f-H_0}
\Bigl[ {\lambda\over {\cal E}_i-H_0} H_1 -H_1{\lambda\over {\cal E}_f-H_0} 
\Bigr] {\lambda\over {\cal E}_i-H_0} H_2 
\Biggr\}\eta_i 
\end{eqnarray}
\narrowtext
By contrast to eqs.~(\ref{2.12}) and (\ref{2.13}), 
eqs.~(\ref{2.28}) and (\ref{2.29}) involve no references to the full energy and
define OBEP and TBEP interactions to definite order.

The unitary transformation method has some significant advantages over the
Tamm-Dancoff approximation:
\begin{itemize}
\item The wave functions are orthonormal, and remain so when $A$
is truncated to a definite order.
\item The effective Hamiltonian is Hermitean and depends only on the 
{\it free} (not full) energy of the observables states. 
\item The interaction is well-defined to any definite order: the definition
of the OBEP interactions does not change when higher-order interactions are 
introduced.
\end{itemize}
From the above discussion we realize that, given any microscopic model 
definition for the NN interaction, a corresponding definition of the 3N 
interaction follows immediately.  However, we have also seen 
that the unitary transformation method possesses operators which are entirely 
absent from the Tamm-Dancoff result.  Without even writing down a Lagrangian,
we will later see that, in applications assuming orthogonal wave functions, 
the Tamm-Dancoff approximation (as used to define the Bonn model) guarantees 
a 3N force which possesses a hard core, whereas the unitary 
transformation method (as used to define the RuhrPot model) ensures the 
corresponding dynamics vanish in the 3N system.

%%%%%%%%%%%%%%%%%%%%%%%%%%%%%%%%%%% Chapter 3 %%%%%%%%%%%%%%%%%%%%%%%%%%%%%%%%%%
\section{OBEP and The Hard Core of the NN-interaction}\label{sec3}
The characteristic behaviour of the NN-interaction has been 
known for a long time. A weak attraction results at long ranges almost entirely 
from the exchange of the bound-state $\pi$-meson. At intermediate ranges
multiple-pion exchange becomes important, so that one boson-exchange 
potentials (OBEPs) generally need to account for correlated 2$\pi$ and 
3$\pi$ exchange by including the $\rho$ and $\omega$ mesons.
The apparently less important uncorrelated 2$\pi$ exchange 
requires explicit calculation of a two-boson exchange potential (TBEP). 
Moving towards the hard core, heavier mesons gain an obvious importance
as a means of effectively describing highly correlated and complicated
exchange processes.

The generic form of the NN interaction is written symbolically as,
\begin{eqnarray}
V_{\rm obep} &=& \sum_{\alpha_\pi =\pi-{\rm like}} V_{\alpha_\pi} 
              =  V_{\pi} + V_{\pi'} + V_{\pi''} + \cdots
\nonumber \\
             &+& \sum_{\alpha_\rho =\rho-{\rm like}} V_{\alpha_\rho} 
              =  V_{\rho} + V_{\rho'} + V_{\rho''} + \cdots
\nonumber \\
             &+& \sum_{\alpha_\omega =\omega-{\rm like}} V_{\alpha_\omega} 
              =  V_{\omega} + V_{\omega'} + V_{\omega''} + \cdots
          \\ &\vdots&
\end{eqnarray}
where, for each $(J^\pi;T)$, an entire spectrum of exchange processes 
contribute to the interaction. The problem is how to include such dynamics
without introducing too many parameters.

The conventional approach in meson physics is simply to forget about everything 
except the mesons with masses under about 1 GeV. This certainly seems reasonable
when we consider the individual contribution of any given heavy meson. After 
all, a large meson mass causes the meson propagator to suppresses the 
contribution to the NN interaction at low energies, and in any event the 
coupling constants and form factor scales are mostly unknown.

But is it reasonable to neglect the {\it summed} contribution of {\it all} of 
the additional exchange processes? For simplicity, let's focus on the $\pi$-like
contributions to OBEP, {\it i.e.}, the part of the NN interaction characterized
by ($J^\pi;T$)=($0^-;1$) exchange with a (not necessarily sharp) mass 
$m_{\alpha_\pi}$.  This does not restrict us to consider only bound-state 
heavy meson-exchanges, but it does imply a common operator structure 
${\cal M}_{\pi-{\rm like}}$, so that the NN interaction takes the form,
\begin{eqnarray}\label{eq3.1}
V_{\pi-{\rm like}} 
&=&  {\cal M}_{\pi-{\rm like}} \sum_{\alpha_\pi =\pi-{\rm like}} 
{ g^2_{{\rm NN}\alpha_\pi} \over 4\pi} 
 {F^2_{{\rm NN}\alpha_\pi} (Q^2)\over m_{\alpha_\pi}^2 + Q^2}
\nonumber \\
&=&  {\cal M}_{\pi-{\rm like}} 
{ g^2_{{\rm NN}\pi} \over 4\pi} {F^2_{{\rm NN}\pi} (Q^2)\over m_\pi^2 + Q^2}
\nonumber \\
&+& {\cal M}_{\pi-{\rm like}}  \sum_{\alpha_\pi \neq \pi} 
{ g^2_{{\rm NN}\alpha_\pi} \over 4\pi} 
 {F^2_{{\rm NN}\alpha_\pi} (Q^2)\over m_{\alpha_\pi}^2 + Q^2}
\end{eqnarray}
where we have separated out the lightest meson contribution from the 
remaining processes. Note that the additional ($J^\pi;T$)=($0^-;1$) exchange
contributions are required by completeness and necessarily add coherently with 
{\it no possibility for cancellation}. 

From eq.~(\ref{eq3.1}) it appears that we have conventional OBEP and a 
summation over heavy meson-exchanges processes. If the summation $\alpha$ 
over the ($J^\pi;T$)=($0^-;1$) exchanges could be truncated at sufficiently 
low order, this interpretation might suffice. If not, then eq.~(\ref{eq3.1}) 
introduces a semi-phenomenological description of all ($J^\pi;T$)=($0^-;1$) 
exchange dynamics - not just meson-exchanges.

In the RuhrPot NN-interaction these additional contributions are retained 
by writing the NN-interaction as,
\begin{eqnarray}
V_{\pi-{\rm like}} 
&\sim&  
{\cal M}_{\pi-{\rm like}} 
{ g^2_{{\rm NN}\pi} \over 4\pi} {F^2_{{\rm NN}\pi} (Q^2)\over m_\pi^2 + Q^2}
\nonumber \\
&+& {\cal M}_{\pi-{\rm like}} 
\Sigma_{\pi-{\rm like}}F^2_{{\rm NN}\pi-{\rm like}} (Q^2)
\end{eqnarray}
where the additional `contact' term introduces an effective constant 
$\Sigma_{\pi-{\rm like}}$ through a closure approximation. 
The RuhrPot NN interaction also includes analogous
$\Sigma_{\rho-{\rm like}}$, $\Sigma_{\omega-{\rm like}}$ and 
$\Sigma_{\epsilon-{\rm like}}$ contact terms.

The contact interactions completely dominate in the region 
of the hard core, but are essentially vanishing at the larger distances where
the meson-exchange dynamics take over. It is important to note that 
the meson-exchange contributions in the RuhrPot model are heavily cut down in 
the region of the hard core. This results because the meson-nucleon vertices 
are dressed with form factors that obey the asymptotic $Q^2$-dependence 
predicted by perturbative QCD \cite{BR81}. In particular, the Dirac and Pauli 
form factors used in the RuhrPot model are given by\cite{GK85}
\begin{mathletters}
\begin{equation}
F^{(1)} = {\Lambda_1^2 \over \Lambda_1^2 + \hat{Q}^2}
            {\Lambda_2^2 \over \Lambda_2^2 + \hat{Q}^2}
\end{equation}
\begin{equation}
F^{(2)} = F^{(1)} 
            {\Lambda_2^2 \over \Lambda_2^2 + \hat{Q}^2}
\end{equation}
\end{mathletters}
where
\begin{equation}
\hat{Q}^2 = Q^2 \log\left[ {\Lambda_2^2+Q^2 \over \Lambda_{\rm QCD}^2 }\right] 
              / \log\left[ {\Lambda_2^2     \over \Lambda_{\rm QCD}^2 }\right] 
\end{equation}
The meson-scales $\Lambda_1$ have been obtained from direct calculation of the 
meson-baryon form factors\cite{KG83,JF95}, and the QCD scales 
$\Lambda_2$ and $\Lambda_{\rm QCD}$ are obtained from a fit to the 
nucleon electromagnetic form factors at $Q^2$ ranging to about 
30~GeV$^2$.

The RuhrPot NN interaction \cite{RP94} fits the scattering phases with 
$\chi^2$/datum=1.6 and the deuteron observables 
$E_D$=-2.224575(9)~MeV, 
$Q_D$=0.2860(15)~fm${}^2$, 
$A_S$=0.8846(8)~fm${}^{1\over 2}$
D/S=0.0272(4) and
$r_D$=1.9560(68)~fm
(none of which are fitted) are all reasonably predicted as
$E_D$=-2.224~MeV, 
$Q_D$=0.276~fm${}^2$, 
$A_S$=0.882~fm${}^{1\over 2}$
D/S=0.025 and
$r_D$=1.932~fm.
The ${\rm SU(3)}_{\rm F}$ result of 
$g^2_{{\rm NN}\omega}$ /$g^2_{{\rm NN}\rho}$ = 9 is retained, which can be 
compared to values of around 27 required in conventional BEP \cite{Bonn}.
This clearly has consequences for a consistent specification of the
$\rho\pi\gamma$ and $\omega\pi\gamma$ exchange currents \cite{JE95a,JE95b}
required in the calculation of electromagnetic observables.
%
%%%%%%%%%%%%%%%%%%%%%%%%%%%%%%%%%%% Chapter 4 %%%%%%%%%%%%%%%%%%%%%%%%%%%%%%%%%%
\section{Contact interactions in the 3N-System}\label{sec4}
In the previous section we recalled how the RuhrPot NN-interaction retains 
contact terms to include the summed ($J^\pi;T$) exchange dynamics. This lead 
naturally to a hard core in the OBEP NN-interaction. But what about the 
3N force? What happens when the TBEP dynamics of the 3N-interaction is extended
to include contact interactions? This simple question needs to be answered with
some care.

In section~\ref{sec2} we obtained two equivalent and model-independent 
definitions of the TBEP. In the {\it Tamm-Dancoff} approximation, as is used 
to define the full Bonn potential for example, 
the wave functions necessarily violate the conventional orthonormality 
requirement and the explicit energy dependence destroys the Hermiticity
of the effective interaction. When the energy of the $A$-particle system is 
assumed to be conserved in all intermediate states the need for non-orthogonal
wave functions remains, but the non-Hermiticity of the effective interaction 
is no longer apparent and the resulting TBEP in the 3-body system is given by,
\begin{mathletters}\label{4.1}
\begin{eqnarray}
V_{\rm eff}^{(4?)} &=& \eta \Biggl\{
H_1 {\lambda\over E-H_0} H_1 {\lambda\over E-H_0} H_1 {\lambda\over E-H_0} H_1 
\\ &+&
H_2 {\lambda\over E-H_0} H_1 {\lambda\over E-H_0} H_1
\\ &+&
H_1 {\lambda\over E-H_0} H_2 {\lambda\over E-H_0} H_1
\\ &+&
H_1 {\lambda\over E-H_0} H_1 {\lambda\over E-H_0} H_2
\Biggr\}\eta 
\end{eqnarray}
\end{mathletters}
where $E$ is the full (including binding) energy of the 3-nucleon system.

Alternatively, from section~\ref{sec2} we recall that a Hermitian interaction
requiring orthonormal wave functions can be obtained from the {\it unitary 
transformation} procedure - as has been done to define the RuhrPot NN- and 
3N-interactions. Adopting energy conservation for comparison with 
eq~(\ref{4.1}), the resulting TBEP contributions to the 3-body system are 
given by,
\begin{mathletters}\label{4.2}
\begin{eqnarray}
V_{\rm eff}^{(4)} &=& \eta \Biggl\{
H_1 {\lambda\over {\cal E}-H_0} H_1 {\lambda\over {\cal E}-H_0} 
H_1 {\lambda\over {\cal E}-H_0} H_1 
\\ &+& H_2 {\lambda\over {\cal E}-H_0} H_1 {\lambda\over {\cal E}-H_0} H_1
\\ &+& H_1 {\lambda\over {\cal E}-H_0} H_2 {\lambda\over {\cal E}-H_0} H_1
\\ &+& H_1 {\lambda\over {\cal E}-H_0} H_1 {\lambda\over {\cal E}-H_0} H_2
\\ &-& {1\over 2} H_1 {\lambda\over {\cal E}-H_0} \Biggl[
  {\lambda\over {\cal E}-H_0} H_1 \eta H_1 
\nonumber\\ && \qquad
+ H_1 \eta H_1 {\lambda\over {\cal E}-H_0} 
\Biggr] {\lambda\over {\cal E}-H_0} H_1 
\Biggr\}\eta 
\end{eqnarray}
\end{mathletters}
where ${\cal E}$ is the free energy of the $3$-body system.

Eq~(\ref{4.2}a) describes meson-recoil, vector-meson decay and baryon resonance 
contributions to the 3N-interaction, whereas
eq~(\ref{4.2}b-d) describes contributions involving less than 4 vertices, 
at least one of which is of second order. 
These operators are already different in the Tamm-Dancoff and unitary 
transformation schemes because of the appearance of full and free-particle
energies respectively. However, the most obvious cost in achieving 
Hermiticity and orthonormality in the unitary transformation scheme is 
found in the explicit appearance of the wave function re-orthonormalization 
contributions of eq~(\ref{4.2}e). These are entirely absent in the 
Tamm-Dancoff scheme.  

So does eq.~(\ref{4.2}) imply that the 3N-interaction has a hard core?
It is important to realize that the 3-body force should include all 
$(J^\pi;T)$-exchange dynamics.  Nothing in eq~(\ref{4.2}), or indeed the
projection formalisms described in section~\ref{sec2}, indicates that we can 
arbitrarily truncate the dynamics to include only the lightest exchange 
processes.  As in the NN-interaction, our task is include these additional
contributions without introducing too many parameters. 

Consider the 3N-interactions involving the exchange of two arbitrary mesons,
say $\alpha$ and $\beta$. Eq~(\ref{4.2}a) and (\ref{4.2}e) involve a product 
of coupling constants $g^2_{{\rm NN}\alpha}g^2_{{\rm NN}\beta}$, or
if an arbitrary baryon resonance N$^*$ is excited, 
$g_{{\rm NN}\alpha}g_{{\rm NN}^*\alpha}g_{{\rm NN}^*\beta}g_{{\rm NN}\beta}$.
Conversely, eqs~(\ref{4.2}b-d) involve
$g_{{\rm NN}\alpha}g_{{\rm NN}\alpha\beta}g_{{\rm NN}\beta}$.
As such, when we sum over all mesons $\alpha$ and/or $\beta$, only the 
contributions from eqs~(\ref{4.2}a) and (\ref{4.2}e) without nucleon resonances 
are certain to form a coherent sum with no possibility for cancellation. This 
identifies the most likely source of a possible hard core in the 3N force.

The situation is, however, fundamentally different to the NN-interaction 
because eq.~(\ref{4.2}) involves a linear combination of meson recoil 
and wave function re-orthonormalization processes and these are of opposite 
sign. In Fig~3 we illustrate these contributions for the exchange of arbitrary 
mesons $\alpha$ and $\beta$. The meson-recoil and re-orthonormalization graphs 
are shown only for the time-ordered topologies
$a^\dagger_\beta(3) a^\dagger_\alpha(2) a_\beta(2) a_\alpha(1)$
and 
$a^\dagger_\beta(3) a_\beta(2) a^\dagger_\alpha(2) a_\alpha(1)$
respectively since all other time orderings can reached through
time reversal and permutations of the nucleon numbers.

Denoting the common operator structure for these 
contributions as ${\cal M}$, the non-relativistic contribution 
to the potential energy from the meson-recoil processes of Fig.~3(a-d) is,
\begin{eqnarray}
&&V^{\rm 3N}_{\rm recoil} = {\cal M}\sum_{\alpha\beta} \Bigl[ 
      {-2\over \omega_\alpha(\omega_\alpha+\omega_\beta)\omega_\beta}
\nonumber \\ &&
    + {-1\over \omega_\alpha(\omega_\alpha+\omega_\beta)\omega_\alpha}
    + {-1\over \omega_\beta (\omega_\alpha+\omega_\beta)\omega_\beta}
    \Bigr]
\end{eqnarray}
and the corresponding contribution from the wave function re-orthonormalization
processes of Fig~3(e-h) is,
\begin{eqnarray}
V^{\rm 3N}_{\rm renorm} &=& -{1\over 2}{\cal M}\sum_{\alpha\beta} \left[
     {-2\over \omega^2_\alpha\omega_\beta}
   + {-2\over \omega_\alpha\omega^2_\beta}
    \right]
\end{eqnarray}
In other words, for the RuhrPot 3N interaction we have,
\begin{equation}
V^{\rm 3N}_{\rm recoil}  +
V^{\rm 3N}_{\rm renorm}  = 0\qquad  \Rightarrow {\rm No~Hard~Core}
\end{equation}
Although this cancellation of wave function re-orthonormalization and
meson recoil terms does {\it not} hold in the NN system, in the 3N system 
it holds for all mesons - regardless of their mass and quantum numbers. 
This result is not dependent on the details of any meson-theoretical model
and it eliminates the most likely source of a hard core in the 3N interaction.
The result is good news for existing definitions of the 3N interaction
\cite{TM1,TM2,TM3,TM4,TM5,TM6,TM7,BR1,BR2,BR3,BR4,BR5} 
and rigorous applications \cite{FG11,CC91,FG05,CP2,CP3,DH93}
because it lends support to the notion that the 3N force can be reasonably
described with only light meson-exchange dynamics. Preliminary 
applications using a consistent definition of the RuhrPot NN \cite{RP94} 
and 3N (see the appendix) interactions have reported \cite{DH95} a 
noteworthy agreement with experiment. In particular, while the triton binding 
energy calculated with the NN-interaction alone gives E$_{\rm B}$=-7.64~MeV, 
when the consistently defined 3N interaction is included the result becomes 
E$_{\rm B}$=-8.34~MeV. This compares favourably with the experimental result of 
E$_{\rm B}$=-8.48~MeV. 

We stress that the unitary transformation procedure described in 
section~\ref{sec2} is central to our definition of the RuhrPot 3N interaction. 
It ensures only Hermitian and energy independent operators arise and that 
they are to be taken between orthonormal wave functions.
It generates wave function re-orthonormalization terms that cancel the recoil 
dynamics, and we have seen that this is central to eliminating the most likely 
source of a hard core in the 3N interaction. 

Had we adopted a procedure like the Tamm-Dancoff approximation, as has been done
to define the full Bonn potential, our results would be changed completely.
From chap~\ref{sec2} we realize that such operators would be non-Hermitian and 
energy dependent and that they would need to be computed between non-orthogonal 
wave functions. Moreover, since there would be no wave function 
re-orthonormalization terms, nothing would cancel the recoil dynamics and 
a hard core in the 3N interaction would result from the summed recoil dynamics.
In other words, for a 3N interaction consistent with the full Bonn potential,
\begin{equation}
V^{\rm 3N}_{\rm recoil} \neq 0\qquad  
\Rightarrow 
{\rm Hard~Core~Exists}
\end{equation}

Both the NN and 3N Tamm-Dancoff effective interactions become consistent with 
the unitary transformation results only when Hermiticity is restored by 
removing the spurious energy dependence and the matrix elements are computed 
with re-orthonormalized wave functions. 
Until such rigor is introduced into 3-body applications that use Tamm-Dancoff 
effective interactions, it is to be hoped that any discrepancies with the
(already rigorous) unitary transformation results will not be interpreted 
in terms of model Lagrangians.
%
%
%%%%%%%%%%%%%%%%%%%%%%%%%%%%%%%%%%% Conclusions %%%%%%%%%%%%%%%%%%%%%%%%%%%%%%%%
\section{Conclusions}\label{sec5}
The dynamics that generates the hard core in RuhrPot NN interaction has been 
considered in the irreducible 3-body force. After presenting a detailed summary 
of the formal definitions of OBEP and TBEP, we recalled that the hard core of 
the NN interaction arises naturally when the $(J^\pi;T)$-exchange dynamics is no
longer truncated to include only the lightest few mesons. We explored the effect
of introducing such dynamics into the 3N system. We isolated those contributions
that necessarily add coherently and therefore cannot possibly be expected to 
cancel. Finally, we showed that these contributions to the irreducible 3N force
vanish identically in the static limit. This is a model independent result.
It lends support to the conventional assumption that the 3N force can reasonably
be described using only light-meson-exchange dynamics.

A completely different result is expected when the
Tamm-Dancoff/Bloch-Horowitz procedure is used to define the NN and 
3N potentials. In that approach, (e.g. full Bonn potential) both the
NN {\it and} 3N potentials contain non-vanishing contributions from the 
coherent sum of meson-recoil dynamics and the possibility of a 
large hard core requiring explicit calculation cannot be ruled out.
\acknowledgements
This work is supported by COSY-KFA J\"ulich (41140512)
and Deutsche Forschungsgemeinschaft (Ga 153/11-4).
%%%%%%%%%%%%%%%%%%%%%%%%%%%%%%%%%%%  Appendix  %%%%%%%%%%%%%%%%%%%%%%%%%%%%%%%%%
\appendix
\section{The RuhrPot 3N interaction}
The RuhrPot 3N interaction is obtained from the unitary transformation
result of eq.~(\ref{2.29}). We present here the leading order 
contributions shown in Fig~1(c-e) in the non-relativistic limit. 
The meson recoil contributions are neglected because they are exactly 
canceled by the wave function re-orthonormalization terms -- the later 
being absent in the Tamm-Dancoff definition of TBEP. 
All coupling constants and form factors are taken to be consistent with
the RuhrPot NN interaction \cite{RP94}.

\widetext
For the excitation of a $\Delta$-resonance mediated by 
the exchange of $\pi$- or $\rho$-mesons we have,
\begin{eqnarray}
% Delta - pi - pi
%
%
%
V_{\rm \Delta\pi\pi} &=&
-{2\over 9}
{ g^2_{{\rm NN}\pi} g^2_{{\rm N}\Delta\pi} \over (2\pi)^6 (2m)^4 }
 {F_{{\rm NN}\pi} (\vec{k}_1^2) F_{{\rm N}\Delta\pi}(\vec{k}_1^2) 
  F_{{\rm NN}\pi} (\vec{k}_2^2) F_{{\rm N}\Delta\pi}(\vec{k}_2^2) 
  \over 
  (\vec{k}_1^2+m_\pi^2)(\vec{k}_2^2+m_\pi^2)(m_\Delta-m) 
} 
\nonumber \\  
&&
(\vec{\sigma}_1.\vec{k}_1) (\vec{\sigma}_2.\vec{k}_2) 
\left[ 4 (\vec{k}_1.\vec{k}_2) (\vec{\tau}_1.\vec{\tau}_2)
  - \vec{\sigma}_3.(\vec{k}_1\times\vec{k}_2) 
    (\vec{\tau}_1\times\vec{\tau}_2).\vec{\tau}_3
\right]
\\  
% Delta - pi - rho 
%
%
V_{\rm \Delta\rho\pi}
&=&
-{2\over 9}
{ g_{{\rm NN}\pi} g_{{\rm N}\Delta\pi} g_{{\rm NN}\rho} g_{{\rm N}\Delta\rho}
\over (2m)^4 (2\pi)^6 }
{ F_{{\rm NN}\pi}(\vec{k}_1^2) F_{{\rm N}\Delta\pi}(\vec{k}_1^2) 
  G^M_{{\rm NN}\rho}(\vec{k}_2^2) G^M_{{\rm N}\Delta\rho}(\vec{k}_2^2)
\over (m_\Delta-m)(\vec{k}_1^2+m_\pi^2)(\vec{k}_2^2+m_\rho^2)}
\nonumber\\ &&\qquad 
\times \Bigl\{ 
4 (\vec{\sigma}_1.\vec{k}_1) \vec{\tau}_1.\vec{\tau}_2 
\left[    (\vec{\sigma}_2.\vec{k}_1) (\vec{k}_2)^2 
         - (\vec{\sigma}_2.\vec{k}_2) (\vec{k}_1.\vec{k}_2)
\right]
\nonumber \\ && \qquad\qquad
-(\vec{\sigma}_1.\vec{k}_1) (\vec{\tau}_1\times\vec{\tau}_2).\vec{\tau}_3 
\left[    (\vec{\sigma}_2\times\vec{\sigma}_3).\vec{k}_1 (\vec{k}_2)^2 
        - (\vec{\sigma}_2.\vec{k}_2) \vec{\sigma}_3.(\vec{k}_1\times \vec{k}_2) 
\right] 
\Bigr\}
\nonumber\\ &&
-{2\over 9} 
{ g_{{\rm NN}\pi} g_{{\rm N}\Delta\pi} g_{{\rm NN}\rho} g_{{\rm N}\Delta\rho}
\over (2m)^4 (2\pi)^6 }
{ F_{{\rm NN}\pi}(\vec{k}_2^2) F_{{\rm N}\Delta\pi}(\vec{k}_2^2) 
  G^M_{{\rm NN}\rho}(\vec{k}_1^2) G^M_{{\rm N}\Delta\rho}(\vec{k}_1^2)
\over (m_\Delta-m)(\vec{k}_1^2+m_\rho^2)(\vec{k}_2^2+m_\pi^2)}
\nonumber\\ && \qquad
\times \Bigl\{ 
4(\vec{\sigma}_2.\vec{k}_2) \vec{\tau}_1.\vec{\tau}_2 
\left[   (\vec{\sigma}_1.\vec{k}_2) (\vec{k}_1)^2 
        - (\vec{\sigma}_1.\vec{k}_1) (\vec{k}_1.\vec{k}_2)
\right] 
\nonumber \\ && \qquad\qquad
+(\vec{\sigma}_2.\vec{k}_2) (\vec{\tau}_1\times\vec{\tau}_2).\vec{\tau}_3 
\left[   (\vec{\sigma}_1\times\vec{\sigma}_3).\vec{k}_2 (\vec{k}_1)^2 
       + (\vec{\sigma}_1.\vec{k}_1) \vec{\sigma}_3.(\vec{k}_1\times \vec{k}_2) 
\right] 
\Bigr\}
\\  
% Delta - rho - rho 
%
%
V_{\rm \Delta\rho\rho}
&=& 
-{2\over 9}
{ g^2_{{\rm NN}\rho} g^2_{{\rm N}\Delta\rho} \over (2m)^4 (2\pi)^6 }
{ G^M_{{\rm NN}\rho}(\vec{k}_1^2) G^M_{{\rm N}\Delta\rho}(\vec{k}_1^2) 
  G^M_{{\rm NN}\rho}(\vec{k}_2^2) G^M_{{\rm N}\Delta\rho}(\vec{k}_2^2)
\over (m_\Delta-m)(\vec{k}_1^2+m_\rho^2)(\vec{k}_2^2+m_\rho^2)}
\nonumber\\ && \qquad 
\times \Bigl\{ 
4\vec{\tau}_1.\vec{\tau}_2 
\Bigl[    (\vec{\sigma}_1.\vec{\sigma}_2) (\vec{k}_1)^2  (\vec{k}_2)^2 
         - (\vec{\sigma}_1.\vec{k}_1) (\vec{\sigma}_2.\vec{k}_1) (\vec{k}_2)^2
\nonumber\\ && \qquad\qquad
  - (\vec{\sigma}_1.\vec{k}_2) (\vec{\sigma}_2.\vec{k}_2) (\vec{k}_1)^2
  + (\vec{\sigma}_1.\vec{k}_1) (\vec{\sigma}_2.\vec{k}_2) (\vec{k}_1.\vec{k}_2)
\Bigr] 
\nonumber\\     && \qquad
-(\vec{\tau}_1\times\vec{\tau}_2).\vec{\tau}_3
\Bigl[(\vec{\sigma}_1\times\vec{\sigma}_2).\vec{\sigma}_3 (\vec{k}_1)^2  
      (\vec{k}_2)^2 
    - (\vec{\sigma}_1.\vec{k}_1) (\vec{\sigma}_2\times\vec{\sigma}_3).\vec{k}_1)
      (\vec{k}_2)^2
\nonumber\\ &&  \qquad\qquad
    + (\vec{\sigma}_2.\vec{k}_2) (\vec{\sigma}_1\times\sigma_3).\vec{k}_2) 
      (\vec{k}_1)^2
    + (\vec{\sigma}_1.\vec{k}_1) (\vec{\sigma}_2.\vec{k}_2) 
      \vec{\sigma}_3.(\vec{k}_1\times\vec{k}_2)
\Bigr] 
\Bigr\}
\end{eqnarray}
For the $\rho\rightleftharpoons\pi\pi$, $\epsilon\rightleftharpoons\pi\pi$ 
and $\omega\rightleftharpoons\pi\rho$ contributions we have
\begin{eqnarray}
V_{\rm \rho\pi\pi} &=& 
{ g^2_{{\rm NN}\pi} g^2_{{\rm NN}\rho} \over (2\pi)^6 2m^3 }
{ F_{{\rm NN}\pi} (\vec{k}_1^2) F_{{\rm NN}\pi} (\vec{k}_2^2) 
  G^M_{{\rm NN}\rho}(\vec{k}_3^2 ) 
  F_{\rho\pi\pi}(\vec{k}_1^2,\vec{k}_2^2,\vec{k}_3^2) 
\over (\vec{k}_1^2+m_\pi^2)(\vec{k}_2^2+m_\pi^2)(\vec{k}_3^2+m_\rho^2) } 
\nonumber \\  
&& \quad
(\vec{\sigma}_1.\vec{k}_1) 
(\vec{\sigma}_2.\vec{k}_2) 
\vec{\sigma}_3.(\vec{k}_1\times\vec{k}_2) 
(\vec{\tau}_1\times\vec{\tau}_2).\vec{\tau}_3
\\
V_{\rm \epsilon\pi\pi} &=& 
{g^2_{{\rm NN}\pi}g_{{\rm NN}\epsilon}f_{\epsilon\pi\pi}\over (2\pi)^6 2m^2}
{ F_{{\rm NN}\pi} (\vec{k}_1^2) F_{{\rm NN}\pi} (\vec{k}_2^2) 
  F_{{\rm NN}\epsilon}(\vec{k}_3^2 ) 
  F_{\epsilon\pi\pi}(\vec{k}_1^2,\vec{k}_2^2,\vec{k}_3^2) 
\over (\vec{k}_1^2+m_\pi^2)(\vec{k}_2^2+m_\pi^2)(\vec{k}_3^2+m_\epsilon^2) } 
\nonumber \\  
&& \quad
(\vec{\sigma}_1.\vec{k}_1) 
(\vec{\sigma}_2.\vec{k}_2) 
(\vec{\tau}_1.\vec{\tau}_2)
\\
V_{\rm \omega\pi\rho} &=& 
{ g_{{\rm NN}\pi}  g_{{\rm NN}\rho} g_{{\rm NN}\omega} g_{\pi\rho\omega} 
\over (2\pi)^6 2m^2 m_\rho }
{ F_{{\rm NN}\pi} (\vec{k}_1^2) 
  G^M_{{\rm NN}\rho}(\vec{k}_2^2 ) 
  F_{{\rm NN}\omega} (\vec{k}_3^2) 
  F_{\rho\pi\omega}(\vec{k}_1^2,\vec{k}_2^2,\vec{k}_3^2) 
\over (\vec{k}_1^2+m_\pi^2)(\vec{k}_2^2+m_\rho^2)(\vec{k}_3^2+m_\omega^2) } 
\nonumber \\  
&& \quad
\Bigl[
  (\vec{\sigma}_2.\vec{k}_1) \vec{k}_2^2
- (\vec{\sigma}_2.\vec{k}_2) (\vec{k}_1.\vec{k}_2)
\Bigr]
 (\vec{\sigma}_1.\vec{k}_1) (\vec{\tau}_1.\vec{\tau}_2)
\nonumber\\ &+& 
{ g_{{\rm NN}\pi}  g_{{\rm NN}\rho} g_{{\rm NN}\omega} g_{\pi\rho\omega} 
\over (2\pi)^6 2m^2 m_\rho }
{ F_{{\rm NN}\pi} (\vec{k}_2^2) 
  G^M_{{\rm NN}\rho}(\vec{k}_1^2 ) 
  F_{{\rm NN}\omega} (\vec{k}_3^2) 
  F_{\rho\pi\omega}(\vec{k}_1^2,\vec{k}_2^2,\vec{k}_3^2) 
\over (\vec{k}_2^2+m_\pi^2)(\vec{k}_1^2+m_\rho^2)(\vec{k}_3^2+m_\omega^2) } 
\nonumber \\  
&& \quad
\Bigl[
  (\vec{\sigma}_1.\vec{k}_2) \vec{k}_1^2
- (\vec{\sigma}_1.\vec{k}_1) (\vec{k}_1.\vec{k}_2)
\Bigr]
 (\vec{\sigma}_2.\vec{k}_2) (\vec{\tau}_1.\vec{\tau}_2)
\end{eqnarray}
and for the $\rho\pi$ terms involving an NN$\pi\rho$ vertex on nucleon 3 
we have,
\begin{eqnarray}
V_{\pi\rho} &=& 
{ g^2_{{\rm NN}\pi}  g^2_{{\rm NN}\rho} \over (2\pi)^6 4m^3 }
{ F_{{\rm NN}\pi}^2 (\vec{k}_1^2) 
  G^M_{{\rm NN}\rho}(\vec{k}_2^2 ) 
  F^{(1)}_{{\rm NN}\rho} (\vec{k}_2^2) 
\over (\vec{k}_1^2+m_\pi^2)(\vec{k}_2^2+m_\rho^2)}
(\vec{\sigma}_1.\vec{k}_1)
(\vec{\sigma}_2\times\vec{\sigma}_3).\vec{k}_2
(\vec{\tau}_1\times\vec{\tau}_2).\vec{\tau}_3
\nonumber \\  
&+& 
{ g^2_{{\rm NN}\pi}  g^2_{{\rm NN}\rho} \over (2\pi)^6 4m^3 }
{ F_{{\rm NN}\pi}^2 (\vec{k}_2^2) 
  G^M_{{\rm NN}\rho}(\vec{k}_1^2 ) 
  F^{(1)}_{{\rm NN}\rho} (\vec{k}_1^2) 
\over (\vec{k}_2^2+m_\pi^2)(\vec{k}_1^2+m_\rho^2)}
(\vec{\sigma}_2.\vec{k}_2)
(\vec{\sigma}_1\times\vec{\sigma}_3).\vec{k}_1
(\vec{\tau}_1\times\vec{\tau}_2).\vec{\tau}_3
\end{eqnarray}
\narrowtext
%%%%%%%%%%%%%%%%%%%%%%%%%%%%%%%%%%% References %%%%%%%%%%%%%%%%%%%%%%%%%%%%%%%%%

%%%%%%%%%%%%%%%%%%%%%%%%%%%%%%% Figure Captions %%%%%%%%%%%%%%%%%%%%%%%%%%%%%%%%
\figure{{\bf Fig 1.} 
The $\pi\pi$, $\pi\rho$ and $\rho\rho$ contributions to the Tucson-Melbourne 
interaction $V_{\rm TM}^{\rm 3N}$ are fixed by subtracting the 
forward-propagating Born amplitudes from $\pi$N-scattering, photo-pion 
production and photon-scattering data.  The RuhrPot interaction 
$V_{\rm RP}^{\rm 3N}$ is calculated from an effective meson-baryon model, 
as in the Brazilian force.  All of these approaches involve only 
light-meson-exchange dynamics, so that it remains to be seen if a hard core 
exists in the 3N force.
}
\figure{{\bf Fig 2.} 
The RuhrPot NN-interaction introduces a closure approximation to incorporate the
additional $(J^\pi;T)$-exchange dynamics that is required by completeness. 
Conventional light-meson-exchange dynamics still contributes to the 
NN-interaction at long distances, but the hard core region is now completely 
dominated by {\it contact} interactions. 
}
\figure{{\bf Fig 3.}
The RuhrPot model adopts the unitary transformation definition of the effective 
3N interaction, so that the recoil (a-d) and wave function 
re-orthonormalization (e-h) contributions cancell in the non-relativistic limit.
The dynamics producing the hard core in the RuhrPot NN interaction therefore 
vanishes in the RuhrPot 3N force.  By contrast, potentials defined within the 
Tamm-Dancoff/Bloch-Horowitz definition (e.g. the full Bonn potential) do not 
include wave function re-orthonormalization contributions, so that the 
surviving recoil dynamics sums coherently to create a 3N force that is 
expected to be large when two or three of the nucleons are at small separations.
}
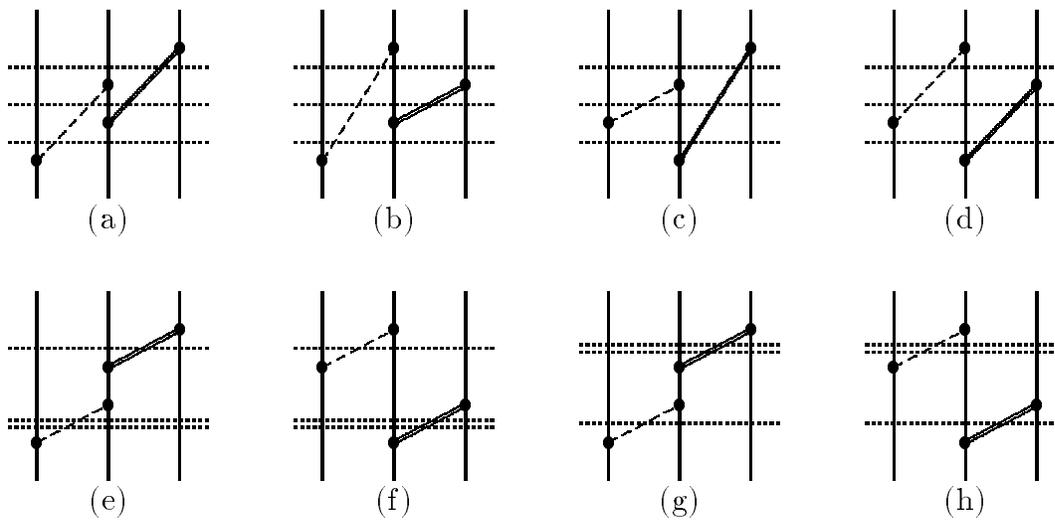
\end{document}